\documentstyle[preprint,aps,epsf]{revtex}
\newcommand{\be}{\begin{eqnarray}}
\newcommand{\ee}{\end{eqnarray}}

\begin{document}
\def\thefootnote{\fnsymbol{footnote}}
\preprint{\begin{minipage}{5cm}
\flushright hep-ph/9605463\\
UAHEP966\end{minipage}}
\vspace*{20mm}
\title{Inclusive Jet $E_T$ Distributions and Light Gluinos}
\author{L. Clavelli\footnote{e-mail: LCLAVELL@UA1VM.UA.EDU}
and I. Terekhov\footnote{e-mail: ITEREKH3@UA1VM.UA.EDU}}
\address{
Department of Physics and Astronomy\\ University of Alabama\\
Tuscaloosa, Alabama 35487\\}
\date{\today}
\maketitle
\begin{abstract}
In the light gluino variant of the minimal supersymmetric model gluino
pairs can be readily produced in collider experiments even if the
squarks are arbitrarily heavy. This enhances the jet transverse energy
distributions. In addition the slower running of the strong coupling
constant in the presence of light gluinos leads to a further
enhancement at higher transverse energies relative to the standard
$QCD$ expectations. Finally, the enhanced squark gluino production
would lead to a Jacobian peak in the $E_T$ distribution at about
$M_{\tilde Q}/2$. These effects are of about the right magnitude to
explain anomalies observed by the $CDF$ and $D0$ collaborations.
\thispagestyle{empty}
\setcounter{page}{0}
\end{abstract}
\pacs{11.30.Pb,14.80.Ly}
\par
Of all the proposals for physics beyond the Standard Model,
Supersymmetry ($SUSY$) seems to be the most theoretically
well-motivated from the aesthetic point of view due to its moderating
of the singular behavior of field theory. In addition there are
successful $SUSY$ unification predictions of the weak angle -- strong
coupling constant correlation and of the $b/\tau$ mass ratio to top
quark mass correlation. Therefore for reasons of economy it is natural
to expect that every deviation from the Standard Model should either
disappear with better statistics or should find its explanation in
terms of $SUSY$. It is generally accepted that current experiments do
not rule out a gluino and photino in the low energy region below $5
GeV$ \cite{UA1}. In fact, if the photino mass lies above the gluino
mass but not above the mass of the gluino-gluon bound state
(glueballino), the region of gluino mass below about $1 GeV$ is
essentially unconstrained by current experiments \cite{Farrar}.
\par
Although the existence of these low energy windows has long been
known, in the last few years there have been many \cite{many}
observances of weak but positive indications of a light gluino from
various Standard Model anomalies.\par Recently both the $CDF$
\cite{CDF} and $D0$ \cite{D0} collaborations have reported anomalies
in the inclusive jet transverse energy distributions at the Tevatron.
In these inclusive measurements each event with $n$ jets satisfying
certain rapidity cuts is binned $n$ times according to the total
transverse energy $E_T$ of each jet. The data as expected is a steeply
falling function of $E_T$ and is most conveniently discussed in terms
of the function
\be
     r(E_T) = { { d\sigma^{DATA} / dE_T }
                   \over{ d\sigma^{QCD} / dE_T}} .
\ee
\par
Since the two experiments use slightly different rapidity cuts the
data do not in principle have to coincide. In addition $r$ is
unfortunately a mixed experimental-theoretical quantity and depends
among other things on the parton distribution functions $(pdf's)$
adopted, on the value of $\alpha_s$ at some reference scale, say
$M_Z$, and on the $QCD$ scale assumed to be appropriate to these
measurements. The experiments use theoretical cross sections
proportional to $\alpha_s(E_T/2)^2$ in lowest order although
theoretical arguments might be made for using the scale $E_T$ or
$2E_T$. This assumption can affect the quantitative results for $r$
but not the qualitative experimental observations which can be
summarized as follows. $CDF$ \cite{CDF} observes values of $r$ below
unity at low $E_T$ followed by a relatively long region where $r$
seems consistent with unity followed by a region of rapid rise. The
$D0$ preliminary 1994/95 data \cite{D0} are consistent with a roughly
constant value of $r \approx 1.2 \pm .07$ in the region $50<E_T<400$
perhaps rising slightly at high $E_T$ with larger errors. It has been
noted \cite{GMRS} that the $CDF$ $r$ values should be renormalized up
by at least $4\%$ to be consistent with the lower values of the strong
coupling constant preferred by deep inelastic data. If one performs
this renormalization and corrects for the slightly different rapidity
cuts in the two experiments \cite{LaiTung}, the $CDF$ and $D0$ data
are consistent at the $1\sigma$ level and both show a systematic
excess of data over theory. According to \cite{GMRS} the $CDF$ results
can not be reconciled with standard $QCD$ by modifying the $pdf$'s
while retaining consistency with constraints from deep inelastic
scattering. Although other authors have searched for alternative
standard $QCD$ effects such as parton double scattering within the
proton
\cite{Dremin} the data remain interesting as a possible observation
of effects beyond the Standard Model and could be evidence for quark
sub-structure or the existence of hitherto unknown partons. An example
of a non-$SUSY$ explanation outside the Standard Model is given by
\cite{CCS}.
\par
However, according to the philosophy discussed in the introduction,
one should first (or at the same time) explore possible $SUSY$ related
explanations. In the currently leading theoretical approach to $SUSY$
in which the squarks and gluinos have masses in the several hundred
$GeV$ to $1 TeV$ region the production of $SUSY$ particles is orders
of magnitude too small to explain the $E_T$ anomaly. In some limited
regions of $E_T$ virtual $SUSY$ effects lead at most to deviations of
several percent from the standard $QCD$ expectations.
\cite{EllisRoss}
\par
In this note, therefore, we explore the scenario where the gluino lies
in the low energy region while the squarks lie in the hundred $GeV$
region. For definiteness we take the gluino mass to be $0.1 GeV$
although our results are not sensitive to the assumed mass. In this
light gluino variant of the minimal $SUSY$ model there are three
effects which can affect the Fermilab experiments at the level of the
observed anomalies.
\par
1.) In the light gluino case the strong coupling constant runs more
slowly than in standard $QCD$. Since in this paper we intend to deal
with lowest order QCD cross sections, we use also the one-loop
renormalization group equations. We do not expect our results to
change qualitatively in higher orders. The one-loop running of the
coupling is defined by the renormalization group behavior
\be
    4\pi {{d\over{d\ln(Q)}}\alpha_s(Q)^{-1}}= - 2 b_3 ,
\ee
where the standard $QCD$ and $SUSY$ coefficients are
\be
     b_3^{QCD} = -11 + 2 n_f/3 ,
\ee
\be
     b_3^{SUSY} = -11 + 2 n_f(1+n_s/2)/3 + 2 n_g .
\ee
\par
Here $n_f$ is taken to be the number of quarks below mass $Q$ (5 or 6
depending on $Q$), $n_s$ is zero or one depending on whether $Q$ is
below or above the (assumed degenerate) squark mass, and $n_g$ is zero
or one depending on whether $Q$ is below or above the gluino mass. In
the light gluino case $n_g$ is always unity for $Q$ in the multi-GeV
region. The result is that, given the value of $\alpha_s$ at some
reference value, say $M_Z$, $\alpha_s$ lies below the standard $QCD$
expectation at lower values of $Q$ and above at higher values of $Q$.
Since the jet cross sections are proportional to second and higher
order powers of the strong coupling constant, the light gluino
prediction would be for $r$ to be below unity at low values of $E_T$
and rising at high values of $E_T$ in qualitative agreement with the
$CDF$ results. The quantitative predictions, which depend on the
assumed scale for the parton scattering, are discussed below.
\par
2.) A second important effect in the light gluino case is the
appearance of extra jets due to gluino pair production. An extra octet
of light elementary particles might a priori be expected to nearly
double the QCD jet cross sections. Since gluino pairs can be produced
via gluon splitting even without intermediate squarks, these pairs
will contribute at lowest (second) order in $\alpha_s$ throughout the
$E_T$ range of the Fermilab experiments. The lowest order parton level
sub-processes are
\be
     G G \rightarrow {\tilde G}{\tilde G} ,\\
     q {\overline q} \rightarrow {\tilde G}{\tilde G} .
\ee
\par
The first process is independent of the squark mass while there is
some squark mass dependence in the second process due to the
possibility of $t$ and $u$ channel squarks. Neglecting the gluino mass
the parton level differential cross sections for gluino pair
production are (from \cite{sigSource})
\be
   \frac{d \sigma(gg\rightarrow{\tilde G}{\tilde G})}{d t} =
\frac{9g_s^4}{64\pi s^2} \left [ \frac{2tu}{s^2}+\frac{u+t}{s}
+\frac{u}{t}+\frac{t}{u} \right ],
\ee
\be
 \frac{d \sigma(q\overline{q}\rightarrow{\tilde G}{\tilde G})}{d t}
 =\frac{g_s^4}{54 \pi s^2} \left [ \frac{9(t^2+u^2)}{2s^2}+\frac{4t^2}
 {(M^2-t)^2}+\frac{9 t^2}{s(t-M^2)}\right ]+(u\leftrightarrow t) ,
\ee
where $M$ is the (assumed $L-R$ degenerate) squark mass. The
transverse energy of each jet is $E_T=\sqrt{ut/s}$.
\par
The relative importance of these processes to the standard $QCD$
$2\rightarrow 2$ sub-processes is easy to estimate by looking at the
90 degree scattering cross sections $(t=u=-s/2)$. Since $QCD$ cross
sections fall rapidly with parton CM energy, for any required value of
$E_T$ the dominant contributions to the cross section will come from
configurations which produce that $E_T$ with minimum parton CM energy.
This is the configuration of 90 degree scattering. One can then
readily estimate an order of $10\%$ enhancement of the inclusive $E_T$
distributions due to gluino pair production neglecting effect 1. For a
quantitative prediction folding in the various $pdf$'s and including
effect 1 we define the lowest order gluino pair production and
standard $QCD$ contributions to the $p \overline{p}$ inclusive jet
distributions dividing out the overall factor of $\alpha_s^2$. That is
\be
   {d{\tilde \sigma}\over{dE_T}} = {1\over{\alpha_s^2}}
   {d\sigma\over{dE_T}}  .
\ee
\par
In this quantity dependence on the $\Lambda_{QCD}$ parameter enters in
only through the small scaling violations in the $pdf$'s. We also
define
\be
 r_\sigma={{{d\tilde\sigma^{SUSY} / dE_T}
 \over{d{\tilde\sigma^{QCD}} / dE_T}}+1}  .
\ee
\par
Here the $SUSY$ cross sections are those of the above gluino pair
production processes and the $QCD$ cross sections are the standard
contributions to $2\rightarrow 2$ scattering. To incorporate the
effect 1 we need the $SUSY$ to $QCD$ ratio of squared couplings.
\be
r_\alpha(Q_1,Q_2) = \left({\alpha_s^{SUSY}(Q_1)
   \over{\alpha_s^{QCD}(Q_2)}}\right)^2 .
\ee
\par
Obviously, in the full supersymmetric theory the $SUSY$ running of
$\alpha_s$ applies to all the $2 \rightarrow 2$ processes. Therefore,
the theoretical prediction for $r$ is
\be
     r(E_T) = r_\sigma r_\alpha .
\ee
\par
It still remains, of course, to choose the scales $Q_1,Q_2$ above.
Since the experiments refer to a theory with $Q=E_T/2$, we should
certainly use this value in the denominator of $r_\alpha$. If the
optimum value of $Q$ is $E_T$ or $2E_T$ as mentioned above this value
should be used in the numerator of $r_\alpha$. This is a theoretical
point which can only be settled in the context of a full higher order
treatment of the inclusive $E_T$ distribution. For definiteness we use
$Q_1=Q_2=E_T/2$ everywhere. In calculating the reduced cross section
ratio $r_\sigma$ we use the CTEQ3L \cite{CTEQ} parton distributions
although the theoretical results which use the $pdf$'s in both the
numerator and denominator are less sensitive to this choice. The
experimentally quoted $r$, on the other hand, depends on the choice of
$pdf$'s only in the denominator and hence is somewhat sensitive to
this choice. Similarly, the theoretical ratios $r_\sigma$ and
$r_\alpha$ are presumably insensitive to inclusion of higher order
effects since these tend to cancel between numerator and denominator.
\par
3.) A final effect that can be discussed in the light gluino case
comes from the parton sub-process
\be
     qG\rightarrow {\tilde q}{\tilde G} ,
\ee
where $q=u,d$.
\par
In the heavy gluino case this cross section is, of course, strongly
suppressed by phase space relative to the light gluino case. Due to
gluino exchange in the $u$ channel, the cross section is strongly
peaked at low energies and forward direction for the primary produced
gluino \cite{TC}. The squark subsequently decays isotropically in its
rest frame into a quark plus gluino. The result is a Jacobian peak in
the inclusive $E_T$ distribution at approximately $M_{\tilde Q}/2$.
Effect 3 is essentially negligible except in this peak region. The
combined predictions of effects 1, 2, and 3 are shown in the solid
lines of fig. 1 for two different values of the mean
up and down squark masses. The standard $QCD$ prediction $r=1$ is
shown in the dashed line. The dot-dashed line roughly constant near
$r=1.06$ shows the behavior of $r_\sigma$ while the dot-dashed line
beginning near $0.8$ and rising above $1.1$ shows that of $r_\alpha$.
Both curves are shown in the case $M_{\tilde Q}=106 GeV$ only.
In this case the $r$ value peaks near $52 GeV$ and rises rapidly
above $200 GeV$ due primarily to effect 1. In the case of a squark of
mass $460 GeV$ the $r$ value peaks at $223 GeV$ and rises less rapidly
above the peak. In this case the rapid rise due to effect 1 would
begin at $E_T=920 GeV$. Below $200 GeV$ the theoretical curves are
insensitive to the squark mass except in the peak region.  The curve
corresponding to the $106 GeV$ mean valence squark mass includes the
supergravity related degeneracy breaking into four peaks with the
predicted overall splitting of about
$20 GeV$.  The data however does not have sufficient resolution to
convincingly resolve these peaks if, indeed, they are preserved after
hadronization.  The splitting at a mean squark mass of $460 GeV$,
predicted to be only about $3 GeV$ overall, is neglected in the
theoretical curve shown.  It does
not seem possible within this scheme to have valence squark spartners
at both $106 GeV$ and $460 GeV$. Therefore within the light gluino
$SUSY$ framework, we would expect that one or more of the two peaks
should disappear with better statistics. From this point of view it is
perhaps significant that the $D0$ data
\cite{D0} show no enhancement in the $225 GeV E_T$ region.
The $D0$ collaboration has not as yet reported results in the region
$E_T<50 GeV $ which would be useful to rule out or confirm a low $E_T$
peak. The normalization and widths of the peaks are, of course,
predicted in supersymmetry given a light gluino and a squark of fixed
mass. In the heavy gluino theory the squark does not have a prominent
two jet decay and hence would lead to a broader peak at lower $E_T$
with a much lower integrated cross section. A squark in the $500 GeV$
region with a two jet decay would also lead to an enhancement at this
mass in the dijet spectrum measured at the Tevatron \cite{TC}. The
$CDF$ dijet data do not rule out a squark in the region below $200
GeV$ since here the peak would be largely submerged in the standard
$QCD$ background.
\par
In summary we have presented the predictions of the light gluino
$SUSY$ theory for the inclusive jet $E_T$ distribution. The predicted
enhancement over the standard $QCD$ expectations agrees roughly in
shape and magnitude with early results from Fermilab. In particular
the theory predicts a dip below $r=1$ in the low $E_T$ region and a
peak near $M_{\tilde Q}/2$. Since we present ratios of $SUSY$ to
Standard $QCD$ predictions we expect that our results will not be
greatly affected by inclusion of higher order perturbative
contributions nor by choices of $pdf$'s. For instance, the next to
leading order corrections are known to increase the standard $QCD$
cross sections by about ten percent \cite{GMRS} and can be expected to
enhance the $SUSY$ cross sections by a comparable amount leading,
therefore, to a much smaller effect on the $r$ ratio. Nevertheless, if
the anomaly persists as further data accumulates, it will be of
interest to pursue refinements of the theory including higher order
contributions and light gluino effects in the $pdf$'s including the
existence of a gluino sea distribution in the proton (which might also
have a bearing on the spin deficit observed in polarized
deep-inelastic scattering). The current inclusive jet $E_T$
experiments have sufficient sensitivity to establish or rule out the
existence of up and down squarks of mass up to at least $600 GeV$ in
association with a light gluino.
\par
In the course of this analysis we profited from discussions with P.W.
Coulter and L. Surguladze at the University of Alabama and with R.
Harris and A. Bhatti of the $CDF$ collaboration. This work was
supported in part by the Department of Energy under grant
$DE-FG02-96ER40967$.

\begin{figure}
\vskip 1cm
\epsfxsize=6in
\epsfysize=3in
\epsfbox{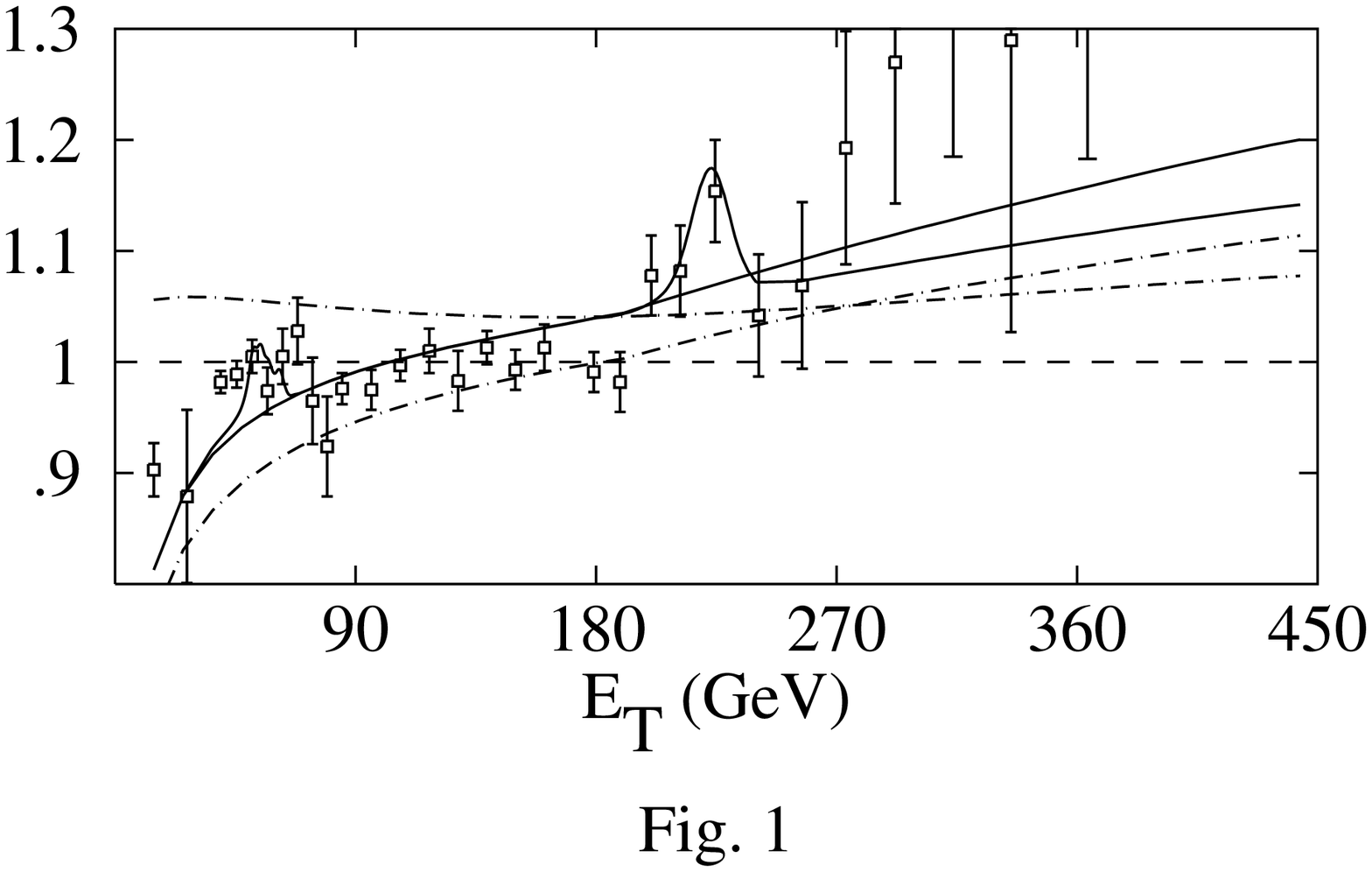}
\vskip 2cm
\caption {
Light gluino predictions for the inclusive
jet $E_T$. The upper and lower dash-dotted curves give the predictions
for $r_\sigma$ and $r_\alpha$ respectively with a squark mass of
$106 GeV$. The solid curves give the combined prediction for $r$ with
an assumed mean squark mass of $460 GeV$ (lower curve at high $E_T$)
or $106 GeV$ (higher curve at high $E_T$). In each case the $r$ value
exhibits a narrow peak near $M_{\tilde Q}/2$. Data from $CDF$ is
superimposed.\label{fig1}}
\end{figure}

\begin{references}
\bibitem{UA1}
C. Albajar et al, UA1 Collab., Phys. Lett. {\bf 198B}, 261 (1987).
\bibitem{Farrar}
G. Farrar, Phys. Rev. {\bf D51}, 3904 (1995).
\bibitem{many}
{L. Clavelli, Phys. Rev. {\bf D46}, 2112 (1992); L. Clavelli, P.W.
Coulter, B. Fenyi, C. Hester, Peter Povinec, and K. Yuan, Phys. Lett.
{\bf B291}, 426 (1992); L. Clavelli, P.W. Coulter, and K. Yuan, Phys.
Rev. {\bf D47}, 1973 (1993); M. Jezabek and J.H. Kuhn, Phys. Lett.
{\bf B301}, 121 (1993); J. Ellis, D. Nanopoulos, and G. Ross, Phys.
Lett. {\bf B305}, 375 (1993); J. Bluemlein and J. Botts, Phys. Lett.
{\bf B325}, 190 (1994); G.R. Farrar, hep-ph/9504295, hep-ph/9508291;
A.L. Kagan, Phys. Rev. {\bf D51}, 6196 (1995); L. Clavelli, Mod. Phys.
Lett. {\bf A10}, 949 (1995); G.R. Farrar and E.W. Kolb, Phys. Rev.
{\bf D53}, 2990 (1996); Peter Povinec, B. Fenyi, and L. Clavelli,
Phys. Rev. {\bf D53}, 4063 (1996).}
\bibitem{CDF}
F. Abe et. al, CDF Collab., Fermilab-Pub-96/020-E (1996).
\bibitem{D0}
D0 Collab., G. Blazey, Proceedings of the XXXI Rencontres de Moriond
(March 1996).
\bibitem{LaiTung}
H.L. Lai and W.K. Tung, hep-ph/9605269, (1996).
\bibitem{GMRS}
E.W.N. Glover, A.D. Martin, R.G. Roberts, and W.J. Stirling,
hep-ph/9603327, (1996).
\bibitem{Dremin}
I.M. Dremin, hep-ph/9604246, (1996).
\bibitem{CCS}
R.S. Chivukula, A.G. Cohen, and E.H. Simmons, hep-ph/9603311, (1996).
\bibitem{EllisRoss}
J. Ellis and D. Ross, hep-ph/9604432, (1996).
\bibitem{sigSource}
R.M. Barnett, H.E. Haber and G.L. Kane, Nucl.Phys. {\bf B267}, 625
(1986); P.R. Harrison and C.H. Llewelyn Smith, Nucl. Phys. {\bf B213},
223 (1983); {\bf E B223}, 542 (1983).
\bibitem{CTEQ}
H.L. Lai et al. (CTEQ) Collaboration, Phys. Rev. {\bf D531}, 4763
(1995).
\bibitem{TC}
I. Terekhov and L. Clavelli, hep-ph/9603390 (1996).
\end{references}
\end{document}